\documentclass{aa}  
\bibpunct{(}{)}{;}{a}{}{,} % to follow the A&A style
\usepackage{graphicx}
\usepackage{subfloat}
\usepackage{txfonts}
\usepackage{hyperref}
\usepackage{xcolor}
\def\ar{ArT\'eMiS }
\def\ars{ArT\'eMiS}
\def\HII{H\,{\sc{ii}}\,}
\def\HI{H\,{\sc{i}}\,}
\def\HIIb{H\,{\sc{ii}}}
\def\mm{$\mu$m\,}
\def\mmb{$\mu$m}
\def\sou{RCW~120 }
\def\sous{RCW~120}

\definecolor{blue}{rgb}{0.0,0.0,1}
\definecolor{green}{rgb}{0.0,1,0.0}
\hypersetup{
    colorlinks,
    linkcolor=blue,
    citecolor=blue,
    urlcolor=blue
    }
\usepackage{etoolbox}
\makeatletter
\patchcmd\@combinedblfloats{\box\@outputbox}{\unvbox\@outputbox}{}{\errmessage{\noexpand patch failed}}
\makeatother

\begin{document} 

   \title{The role of Galactic \HII regions in the formation of filaments}

   \subtitle{High-resolution submilimeter imaging of RCW~120 with ArT\'eMiS\thanks{This publication is based on data acquired with the Atacama
Pathfinder Experiment (APEX) in Onsala program O-0100.F-9305. APEX is a collaboration between the Max-Planck-Institut für Radioastronomie, the European Southern Observatory, and the Onsala Space Observatory.}}

   \author{A. Zavagno\inst{1,2}
          \and
          Ph. Andr\'e\inst{3}
          \and
          F. Schuller\inst{3,4}
          \and
        N. Peretto\inst{5}
        \and
        Y. Shimajiri\inst{3,6}
        \and
        D. Arzoumanian\inst{7}
        \and 
      T. Csengeri\inst{8}
     \and 
     M. Figueira\inst{9}
     \and
     G.
     A. Fuller\inst{10}
     \and
     V. Könyves\inst{11} 
     \and 
     A. Men'shchikov\inst{3}
     \and 
    P. Palmeirim\inst{7}
    \and
     H. Roussel\inst{12} 
     \and
        D. Russeil\inst{1}
      \and
     N. Schneider\inst{8,13}
    \and
    S. Zhang\inst{1}
          }

   \institute{Aix Marseille Univ, CNRS, CNES, LAM, Marseille, France 
                \and 
        Institut Universitaire de France  
           \and
             Laboratoire d'Astrophysique (AIM), CEA/DRF, CNRS, Universit\'e Paris-Saclay, Universit\'e Paris Diderot, Sorbonne Paris-Cit\'e, 91191 Gif-sur-Yvette, France  
             \and
            Leibniz-Institut f\"ur Astrophysik Potsdam (AIP), An der Sternwarte 16, 14482 Potsdam, Germany 
            \and
        School of Physics \& Astronomy, Cardiff University, Cardiff CF24 3AA, UK
            \and
        Department of Physics and Astronomy, Graduate School of Science and Engineering, Kagoshima University, 1-21-35 Korimoto, Kagoshima, Kagoshima 890-0065, Japan 
        \and
           Instituto de Astrof\'isica e Ci{\^e}ncias do Espa\c{c}o, Universidade do Porto, CAUP, Rua das Estrelas, PT4150-762 Porto, Portugal 
               \and
       Laboratoire d'Astrophysique de Bordeaux, Univ. Bordeaux, CNRS, B18N, all\'ee Geoffroy Saint-Hilaire, 33615 Pessac, France
       \and
             National Centre for Nuclear Research, Ul. Pasteura 7, 02-093, Warsaw, Poland 
       \and 
       Jodrell Bank Centre for Astrophysics, Department of Physics and Astronomy, The University of Manchester, Oxford Road, Manchester M13 9PL, UK 
    \and
    Jeremiah Horrocks Institute, University of Central Lancashire, Preston PR1 2HE, UK
    \and 
      Institut d’Astrophysique de Paris, Sorbonne Université, CNRS (UMR7095), 75014 Paris, France    
                \and
            I. Physik. Institut, University of Cologne, 50937, Cologne, Germany   
        %     \email{annie.zavagno@lam.fr}
             }
   \date{Received February 25, 2020; accepted April 3, 2020 }

% \abstract{}{}{}{}{} 
% 5 {} token are mandatory
 
  \abstract
  % context heading (optional)
  % {} leave it empty if necessary  
   {Massive stars and their associated ionized (\HIIb) regions could play a key role in the formation and evolution of filaments that host star formation. However, the properties of filaments that interact with \HII regions are still poorly known.}
  % aims heading (mandatory)
   {To investigate the impact of \HII regions on the formation of filaments, we imaged the Galactic \HII region RCW~120 and its surroundings where active star formation takes place and where the role of ionization feedback on the star formation process has already been studied.}
  % methods heading (mandatory)
   {We used the large-format bolometer camera \ar on the APEX telescope and combined the high-resolution \ar 
data at 350 $\mu$m and 450 \mm with \emph{Herschel}-SPIRE/HOBYS data at 350 and 500 $\mu$m to ensure good sensitivity to a broad range of spatial scales. This allowed us to study the dense gas distribution around RCW~120 with a resolution of 8$\arcsec$ or 0.05~pc at a distance of 1.34~kpc.}
  % results heading (mandatory)
   {Our study allows us to trace the median radial intensity profile of the dense shell of \sous. This profile is asymmetric, indicating a clear compression from the \HII region on the inner part of the shell. The profile is observed to be similarly asymmetric on both lateral sides of the shell, indicating a homogeneous compression over the surface. On the contrary, the profile analysis of a radial filament associated with the shell, but located outside of it, reveals a symmetric profile, suggesting that the compression from the ionized region is limited to the dense shell. The mean intensity profile of the internal part of the shell is well fitted by a Plummer-like profile with a deconvolved Gaussian full width at half maximum (FWHM) of 0.09~pc, as observed for filaments in low-mass star-forming regions. }
  % conclusions heading (optional), leave it empty if necessary 
   {Using \ar data combined with $\emph{Herschel}$-SPIRE data, we found evidence for compression from the inner part of the \sou ionized region on the surrounding dense shell. This compression is accompanied with a significant (factor 5) increase of the local column density. This study suggests that compression exerted by \HII regions may play a key role in the formation of filaments and may further act on their hosted star formation. \ar data also suggest that \sou might be a 3D ring, rather than a spherical structure.}


   \maketitle
%
%-------------------------------------------------------------------
\section{Introduction}
Although the existence of filaments has been known for a long time \citep{sch79,ung87}, the tight relation between filaments and star formation has only been clearly revealed by \emph{Herschel} observations, which introduce a new paradigm for Galactic star formation \citep{and14,mol14}.  Filaments are ubiquitous in the Galactic interstellar medium (ISM) and their properties have been extensively studied and discussed \citep{arz11,pal13,sch14,cox16,li16,sch20}. One of their most striking properties is their typical inner width of 0.1~pc that seems to be quasi-universal \citep{and16,arz19} and which could result from their formation process \citep{fed16,and17}. However, this point is still debated and high-resolution observations are needed to solve the question of filament width universality \citep{hen17,oss19,xu19}. The study of filaments has attracted much attention being the original matter
and host zone of star formation. In particular, the mass per unit length of filaments, which could determine the mass of the cores that will be formed \citep{and19} together with the filaments' ability to fuel the surrounding material toward dense cores allowing for a continuous accretion  \citep{gom18}, can change our view of the star formation process.  Understanding filaments as a whole (formation, evolution) has become a new quest of star formation studies \citep[e.g.,][]{zha19}. 

Another debated question is the role of expanding \HI shell and ionized (\HIIb) regions in the formation of filaments and their influence on star formation. Many observational studies have shown that high-mass star formation is observed at the edges of Galactic \citep{deh05,deh10,tho12} and extragalactic \citep{ber16} \HII regions. This phenomenon is important because at least 30\% of the Galactic high-mass star formation is observed at the edges of \HII regions \citep{pal17}, suggesting that these regions could create the needed conditions for the formation of high-mass stars and might even increase its efficiency \citep{mot18}. Models show the complexity of \HII regions' impact on the surrounding medium \citep{gee17,gee20}.\ However, observational studies demonstrate the importance of ionization feedback in assembling denser material in their photodissociation region (PDR), leading to favorable conditions there for the formation of a new generation of (high-mass) stars \citep{tre14,ork17}. \HII regions have also been proposed as a key agent of the “bubble-dominated” scenario for the formation of filamentary molecular clouds proposed by \citet{inu15}: Previous generations of massive stars generate expanding shells and bubbles; additionally, dense molecular filaments result from multiple large-scale supersonic compressions at the surface of such expanding bubbles and/or at the interface between two bubbles. Three-dimensional magnetohydrodynamic simulations of interacting supershells show filamentary structures at the surface of the shells \citep{nto17, daw15}.  Recent modeling and observations of PDR in high-mass star-forming regions \citep{bro19,goi19b,goi19} also show the importance of understanding the role of ionization feedback on star formation and the role  of \HII regions in the formation of filaments. However, the link between ionization feedback, the possible filamentary structure of PDRs, and star formation observed at the edges of \HII regions is still poorly known.  Addressing this question requires studying the dense PDRs that surround \HII regions and their role in star formation.  
 
RCW~120 is a galactic \HII region located above the galactic plane at $l=348.24^{\circ}$, $b=00.46{^\circ}$ (RA (J2000)=$258.10^{\circ}$, DEC (J2000)=$-38.46^{\circ}$). The region is ionized by a single O6-8V/III star \citep{mar10}. Due to its simple ovoid morphology and nearby location, 1.34~kpc \citep{deh09}, it has been extensively studied in the recent past, especially in order to discuss the properties of star formation observed at its edges \citep{deh09,zav10,fig17,fig18} and also to propose a scenario for its original formation by a cloud-cloud collision \citep{tor15} as well as its kinematics \citep{and15,san18,koh18,kir19}. The wealth of existing data on this region makes it ideal to study the impact of ionization on the structure of its dense PDR and its relation with the observed young star formation.   

With the aim of characterizing the properties of dense PDRs around \HII regions and their relation with the star formation observed at their edges, we present the results of dust continuum mapping observations of the RCW~120 region with the ArTéMiS bolometer camera on the APEX 12 m telescope. The 8$\arcsec$ resolution of ArT\'eMiS at 350 $\mu$m, corresponding to 0.05 pc at the distance of RCW~120, allowed us to reveal the structure of the dense PDR, traced here by a filament, that surrounds the ionized region and to resolve, for the first time, its width. Section~\ref{obs} presents the observations and data reduction. Section~\ref{res} presents the mapping results, which are then discussed in Sect.~\ref{dis}. Conclusions are summarized in Section~\ref{conc}. 

%--------------------------------------------------------------------
\section{ArT\'eMiS observations and data reduction} \label{obs}
Observations of \sou were obtained at 350 and 450 \mm on August 27 and October 10 in 2017 and on May 6, 21, 22, and June 2 in 2018 with the \ar camera on the Atacama Pathfinder Experiment (APEX) (APEX Service Mode run 0100.F-9305(A)). The total on-source observing time is 12.5~hours. \ars\footnote{See http://www.apex-telescope.org/instruments/pi/artemis/ ArTéMiS stands for “ARchitectures de bolomètres pour des TElescopes à grand champ de vue dans le domaine sub-MIllimétrique au Sol” in French.} is a large-format bolometer array camera, which was built by CEA/Saclay and installed in the Cassegrain cabin of APEX, which has a total of 4608 pixels observing at 350 \mm and 450 \mm simultaneously. A complete description of the instrument is given in \citet{and16}.  
We obtained the 350 and 450 \mm maps by coadding the 31 individual maps, obtained at each wavelength, on \sou using the total-power on-the-fly scanning mode with a scanning speed of 30$^{\prime\prime}$/s. The absolute calibration uncertainty is about 30\%. The data reduction procedure is the same as the one described in \citet{and16}. The root mean square (rms) noise is $\sigma \simeq $ 0.25~Jy/8$^{\prime\prime}$-beam at 350 \mm and 0.21~Jy/10$^{\prime\prime}$-beam at 450 \mmb.   

\section{Results} \label{res}
\subsection{Combination of \ar and $Herschel$ data} 
\begin{figure*}[h!]
\includegraphics[width=10cm]{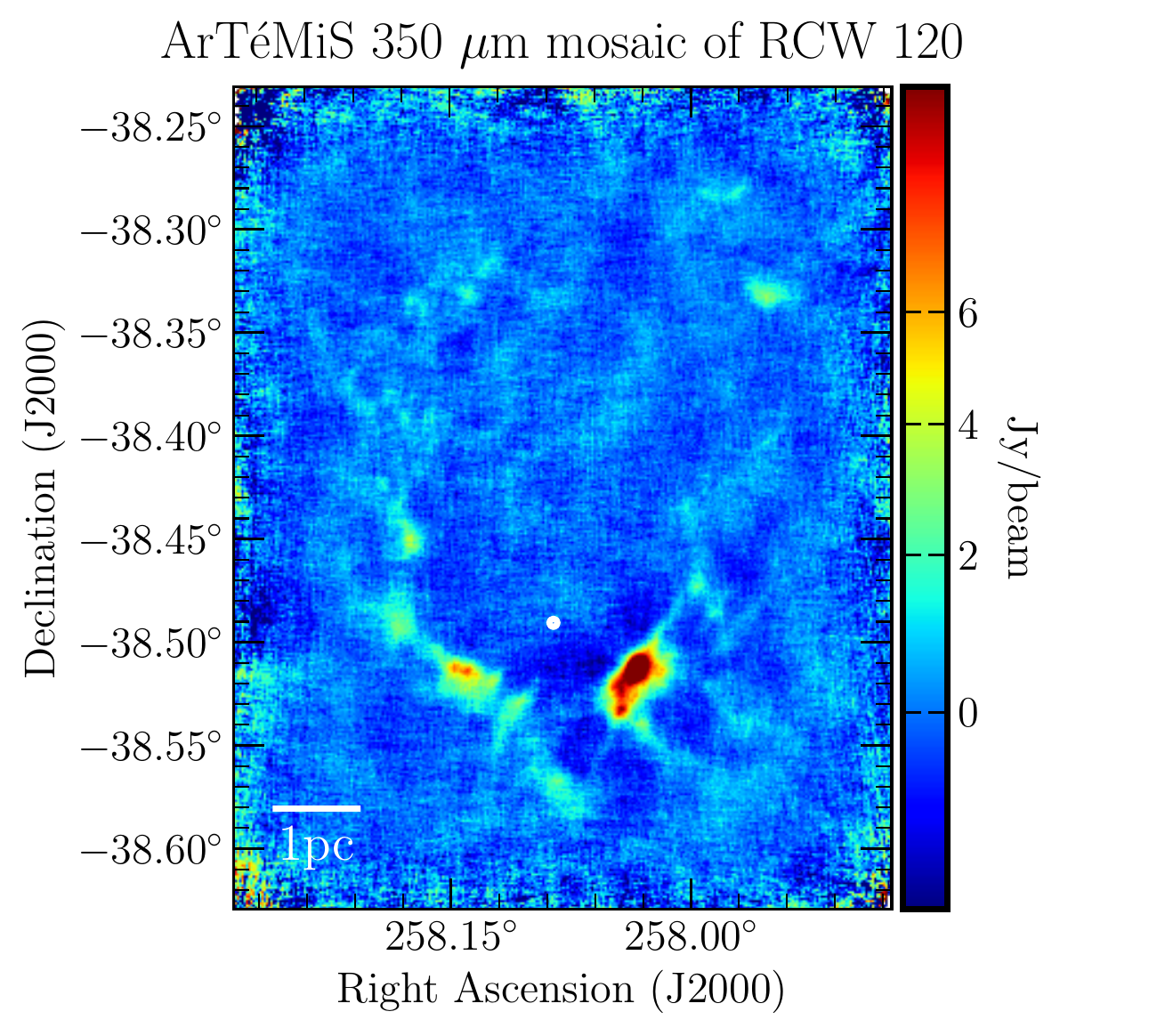}\hfill
\includegraphics[width=10cm]{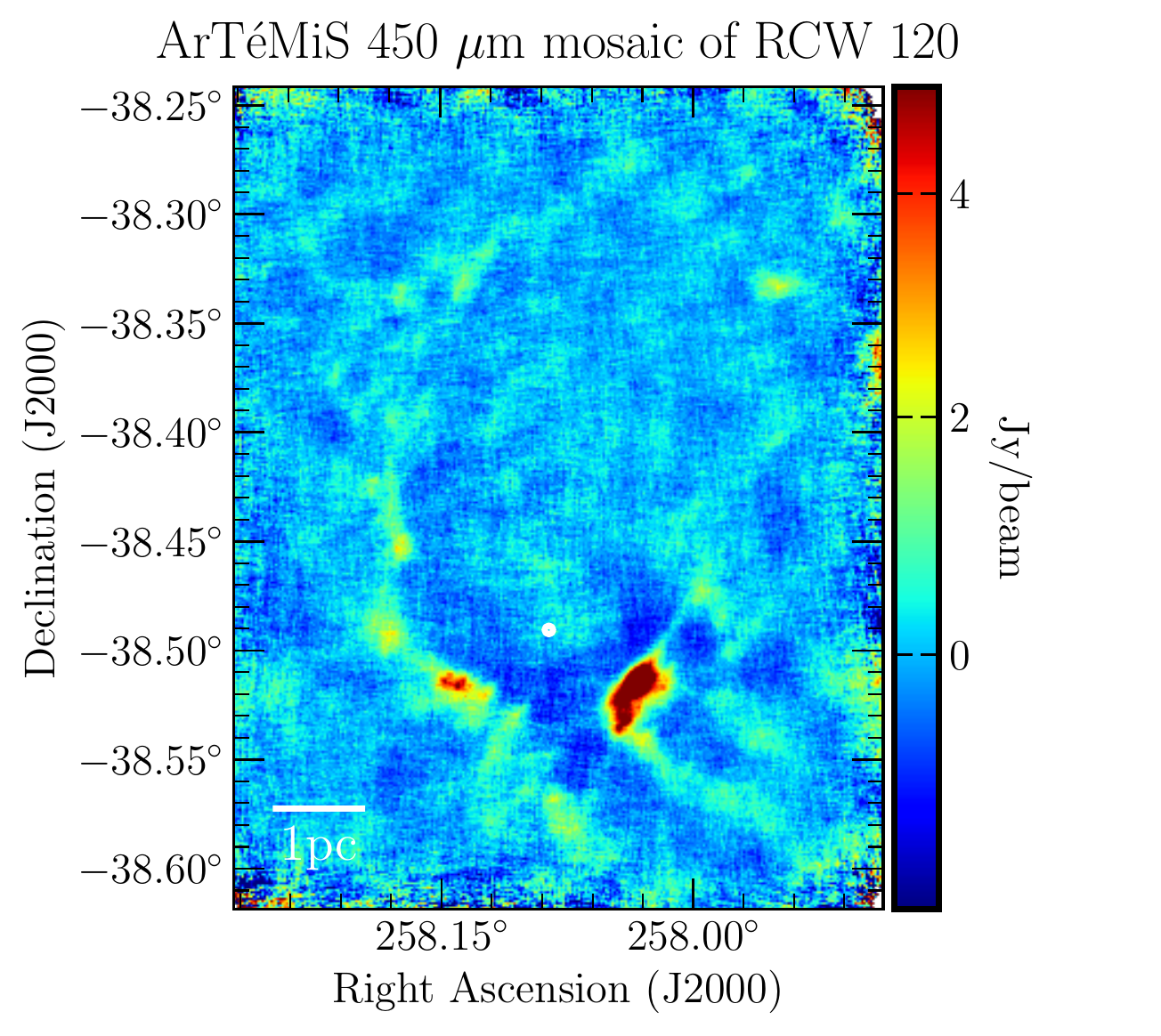}
\caption{\ars-only maps of \sou at 350 \mm (left) and 450 \mm (right) emission observed with Art\'eMiS. The white circle shows the location of the ionizing star (Right ascension (J2000)=258.085833$^\circ$, Declination (J2000)=$-$38.490555$^\circ$). North is up, and east is left. }
\label{map}
\end{figure*}

The \ar raw data are affected by a high level of sky noise, which is strongly correlated over the detectors of the focal plane. Subtracting this correlated sky noise during data reduction makes the resulting image insensitive to angular scales larger than the instantaneous field of view of the camera ($\simeq$ 2$^{\prime} \times$ 4$^{\prime}$). To restore the missing large-scale information, we combined the ArT\'eMiS data with the SPIRE 350 $\mu$m and 500 \mm data from the $Herschel$ HOBYS key project \citep{mot10}, which employed a similar technique as the one used in combining millimeter interferometer observations with single-dish data. The task ``immerge" in the Miriad software package is used \citep{sau95}. A complete description of the method is given in \citet{and16}. Figure~\ref{map} shows the distribution of the \ars-only 350 \mm (left) and 450 \mm (right) emissions observed with \ars. 
%-------------------------------------- Two column figure 
%Fig1
These emissions show the main bright condensations identified previously using continuum millimeter observations \citep{zav07,deh09} that host star formation, which were studied in detail with \emph{Herschel} \citep{zav10,fig17} and ALMA \citep{fig18}. The northern part of the dense PDR layer surrounding the \HII region is seen with fainter emission that delineates the ovoid shape of the region.  

\subsection{Filaments extraction and distribution}
\begin{figure*}[h!]
   \centering
  \includegraphics{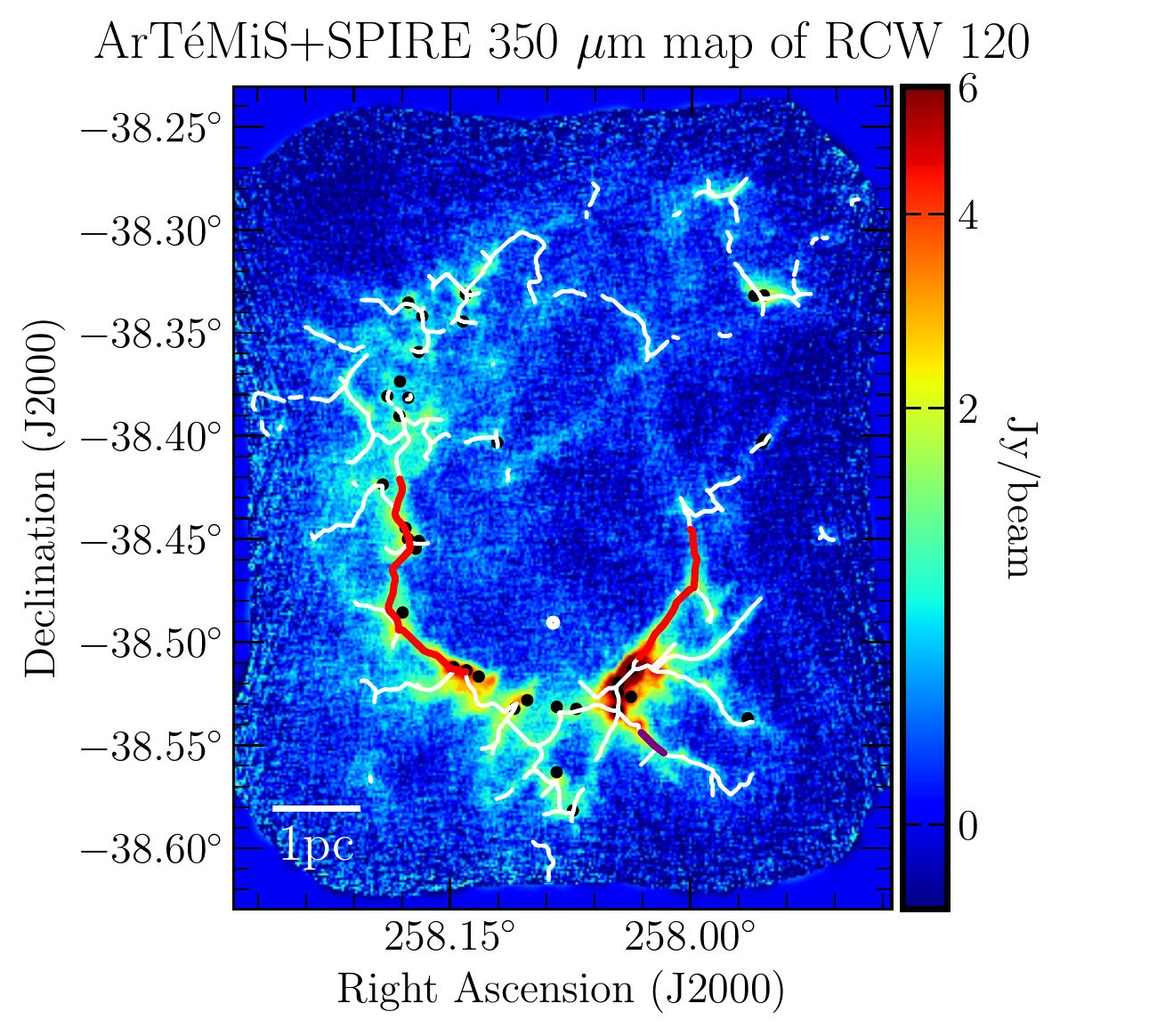}
   \caption{Combined \ars+\emph{Herschel}-SPIRE 350 \mm map of \sous, with DisPerSe skeleton overlaid in white and \emph{Herschel} young sources (solid black circles). The white circle shows the location of the ionizing star. The red parts underline the eastern and western lateral side of the dense shell together with the purple radial filament, which have also been analyzed as separate parts (see text).}
              \label{fs}%
    \end{figure*}

We applied the DisPerSe \citep{sou11} and {\it{getfilaments}} \citep{men13} algorithms to determine the location of the crest of filaments in RCW~120. Both algorithms were applied to the combined $Herschel$ + \ar 350~\mm image. The main crests tracing the dense PDR are shown as the white solid curves in Figure~\ref{fs} together with the young sources identified with \emph{Herschel} and discussed in \citet{fig17}. The sources are distributed along the crests with a higher density of sources in the densest part of the shell, which are observed to the south. Sources are also observed to the north in tight relation with filaments' crests. In this paper, we discuss the filaments' profile over the dense shell. The properties of the filaments and their relation with the observed star formation will be discussed in a forthcoming paper.
%Fig2

\subsection{Radial intensity profile of the shell} \label{rps}
\begin{figure*}[h!]
\includegraphics[width=9cm]{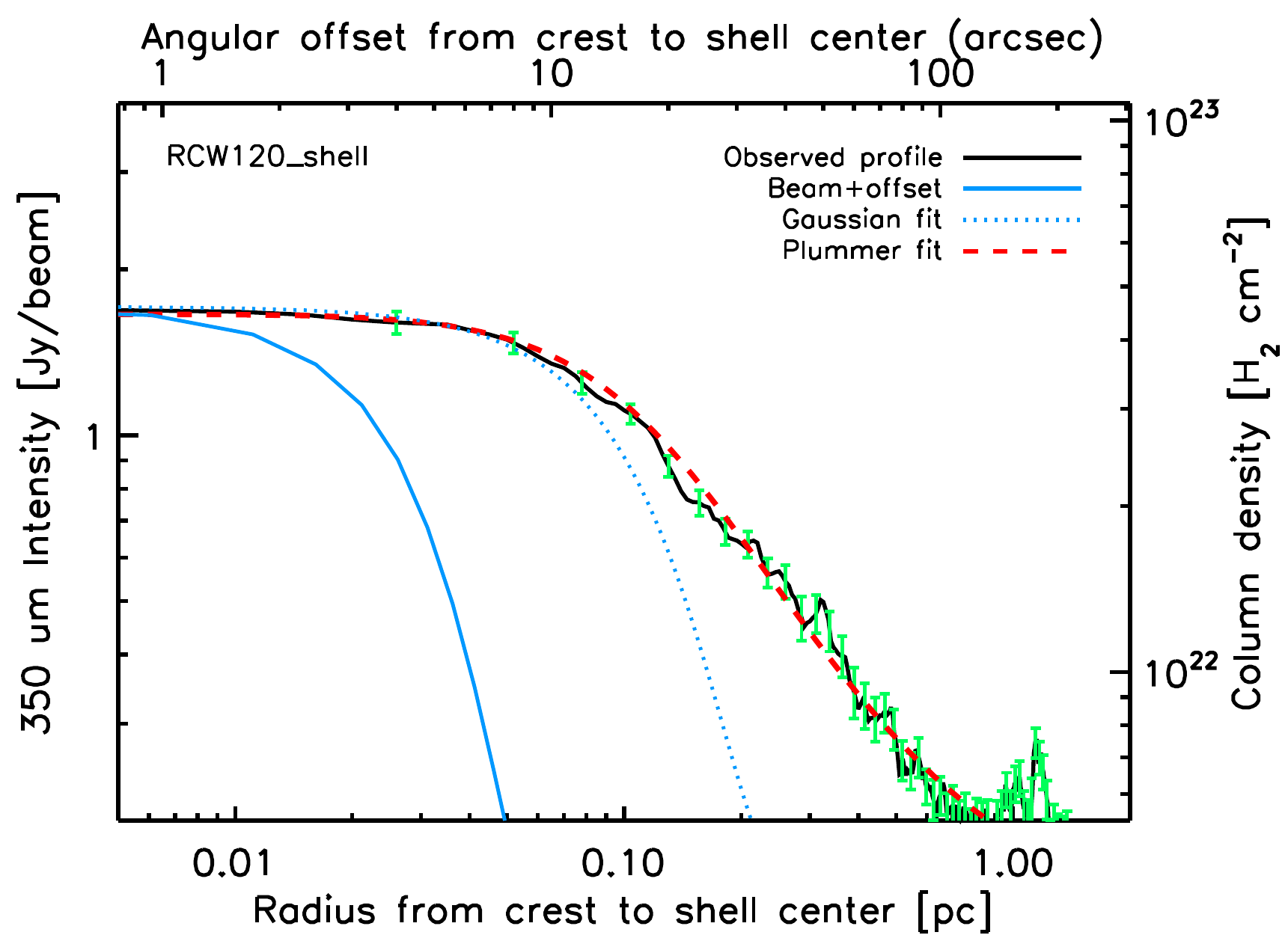}\hfill
\includegraphics[width=9cm]{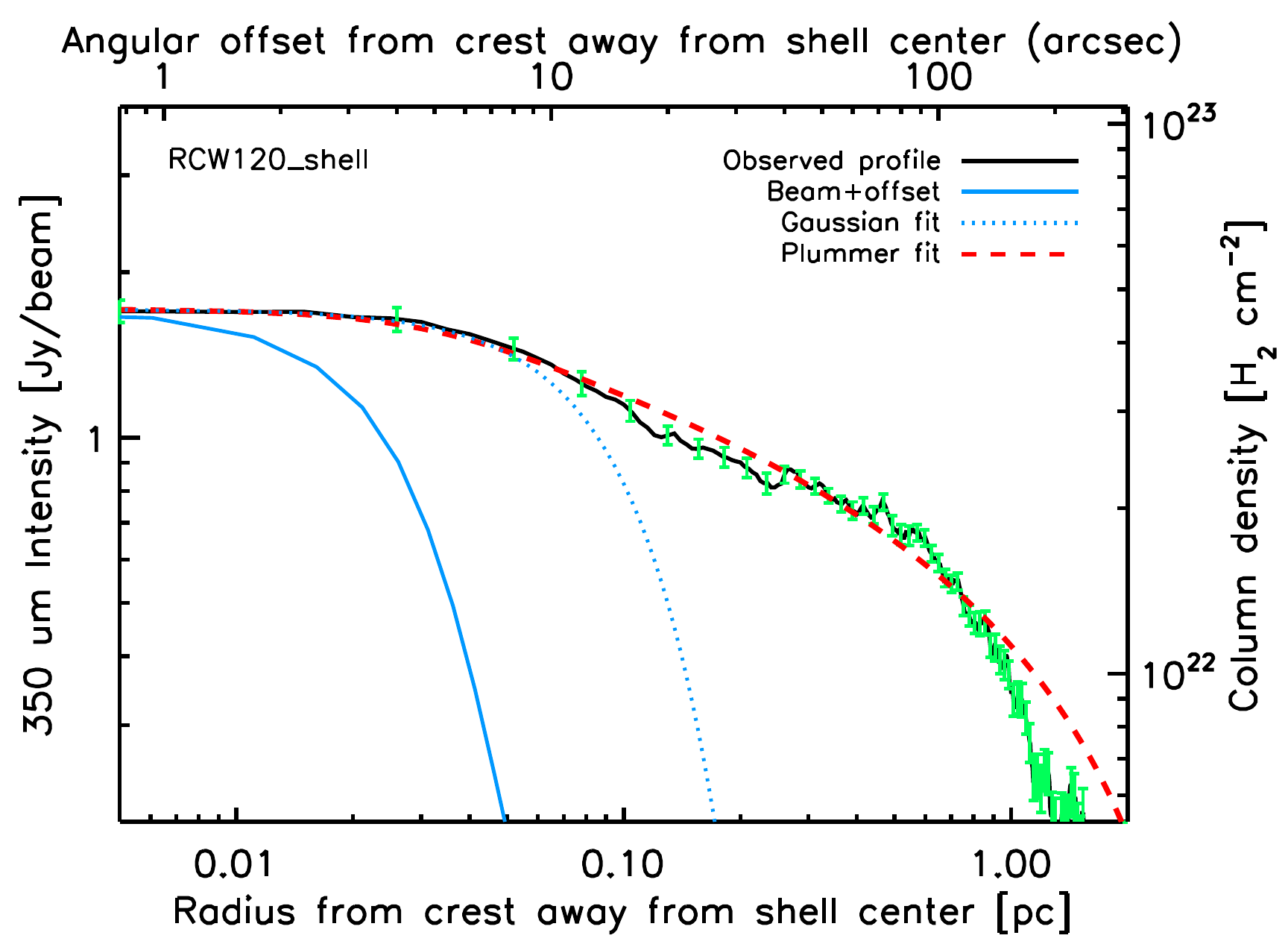}
   \caption{Median radial intensity profile (black curve) of the \sous 's shell measured perpendicular to the crest, toward the shell's center (left) and toward the external part of the shell (right). The green error bars show the standard deviation of the mean intensity profile  (same color coding for the other figures). The blue solid curve shows the effective beam profile of the ArT\'eMiS 350 \mm data as measured on Mars. The blue dotted curve shows the best-fit Gaussian model to the observed profiles. The red dashed curve shows the best-fit Plummer model convolved with the beam. }
              \label{radpro}%
    \end{figure*}

Here, we analyze the profile on the dense shell that surrounds \sous's ionized region and that corresponds to the densest part of \sous's PDR. By taking perpendicular cuts at each pixel along the crest, we constructed a median radial intensity profile. Figure~\ref{radpro} shows the shape of the radial intensity profile toward the internal and external part of the dense shell.  

\begin{figure}[h!]
   \centering
  \includegraphics[scale=0.5]{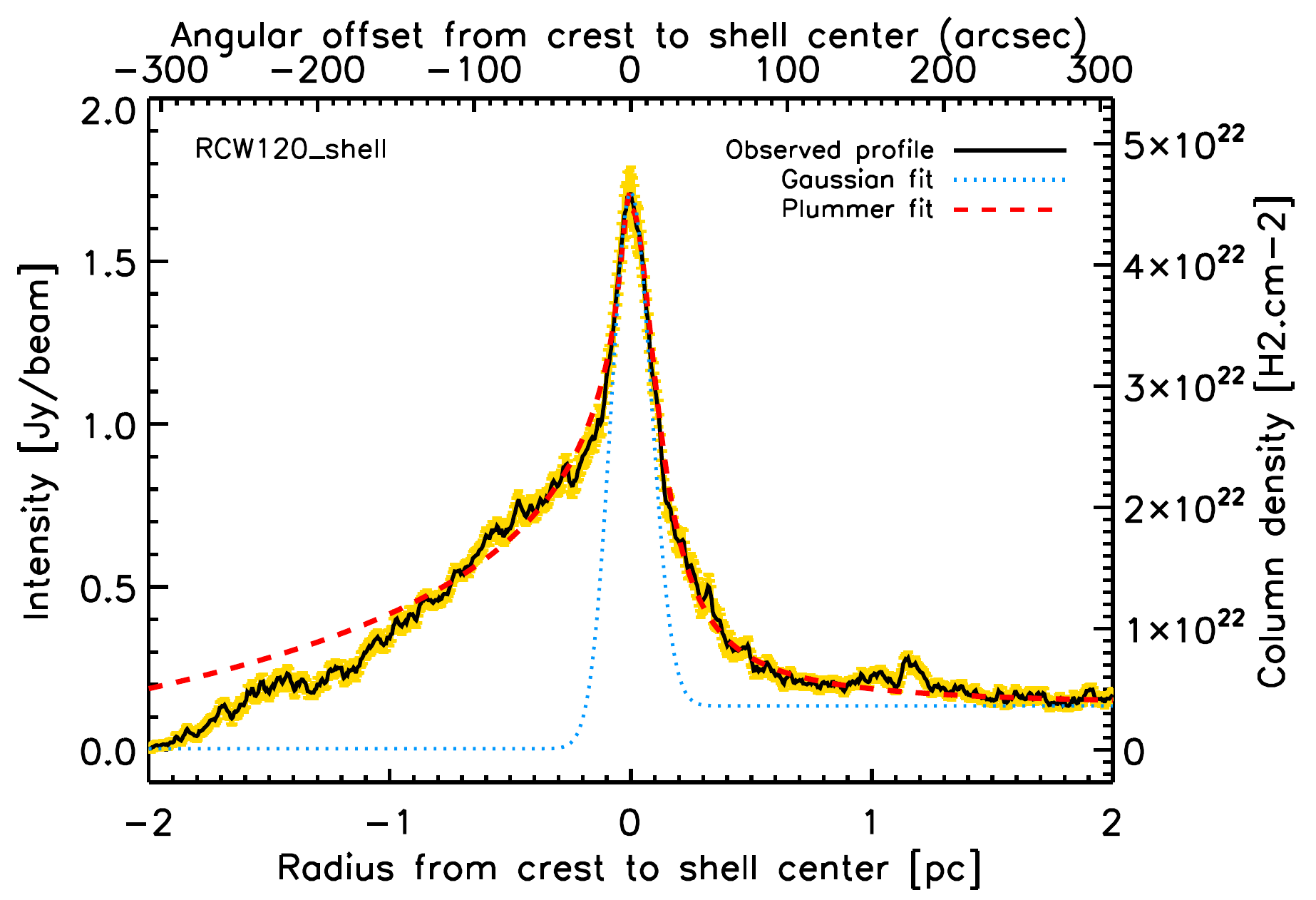}
   \caption{Observed radial column density (and flux density) profile (black curve) of the RCW~120 shell. The Gaussian fit (blue dotted curve) and the Plummer fit (red dashed curve) are shown. The yellow error bars show the 
$\pm 1\sigma$ dispersion of the distribution of radial intensity profile observed along the filament crest  (same color coding for the other figures).  }
              \label{fa3}%
    \end{figure}

We fit all of the observed radial profiles presented in this paper, $I(r)$ as a function of radius $r$ with both a simple Gaussian model, 
\begin{equation}
I_{\rm{G}}(r) =  I_{\rm{0}} \times \exp[-4\ln2 \times (r/FWHM)^2] + I_{\rm{Bg}} 
\end{equation}
\noindent and a Plummer-like model function of the form, 
\begin{equation}
I_{\rm{P}}(r) =  \frac{ I_{\rm{0}}}{[1+(r/R_{\rm{Flat}})^2]^{\frac{p-1}{2}}} + I_{\rm{Bg}} 
\end{equation}
where $I_{\rm{0}}$ is the central peak intensity, $FWHM$ is the physical full width at half maximum of the Gaussian model, $R_{\rm{Flat}}$ represents the characteristic radius
of the flat inner part of the Plummer model profile, $p > 1$ is the power-law index of the underlying density profile, and  $I_{\rm{Bg}}$ is a constant background level. These forms 
were convolved with the approximately Gaussian beam of the \ar data ($FWHM \simeq  8^{\prime\prime}$) before we compared them with the observed profile. 
We assume a dust temperature of $T_{\rm{d}}$ = 15~K (median dust temperature over the region) and a dust opacity law of ${\kappa}_{{\lambda}} = 0.1 \times (\lambda/300$ \mm $)^{{-{\beta}}}$ cm$^2$~per~g (of gas and dust) with an emissivity index of $\beta$=2 \citep{hil83}, which is the same opacity law as the one adopted in the HGBS and HOBYS papers \citep{roy14}.
%Fig.3
Figure~\ref{fa3} shows the linear profile of a median cut through the dense shell; the positive offset is toward the shell center.
This profile shows clear asymmetry, suggestive of a compression from the internal part of the shell, which is likely of a radiative origin. Indeed, \citet{mar10} show that the ionizing star of RCW~120 has a low stellar wind activity. This means that the ionized bubble has a radiative origin and is not created by stellar winds. For the internal part of the shell (compressed part), the mean intensity profile is well fitted by a classic Plummer-like profile \citep{cas86,whi01,arz11,pal13} with $p\simeq$~2 and D$_{Flat}$ $\simeq 0.08 \pm 0.02$ pc, which corresponds to a Gaussian FWHM of $\simeq 0.09 \pm 0.02$ pc. Toward the external part of the shell, the Gaussian FWHM is $\simeq 0.1-0.2$ pc, but the profile is much flatter. The Plummer profile is better fitted with $p \simeq 1$.  

\subsection{Radial profile of the eastern and western lateral sides of the shell}
\begin{figure*}[h!]
\includegraphics[width=9cm]{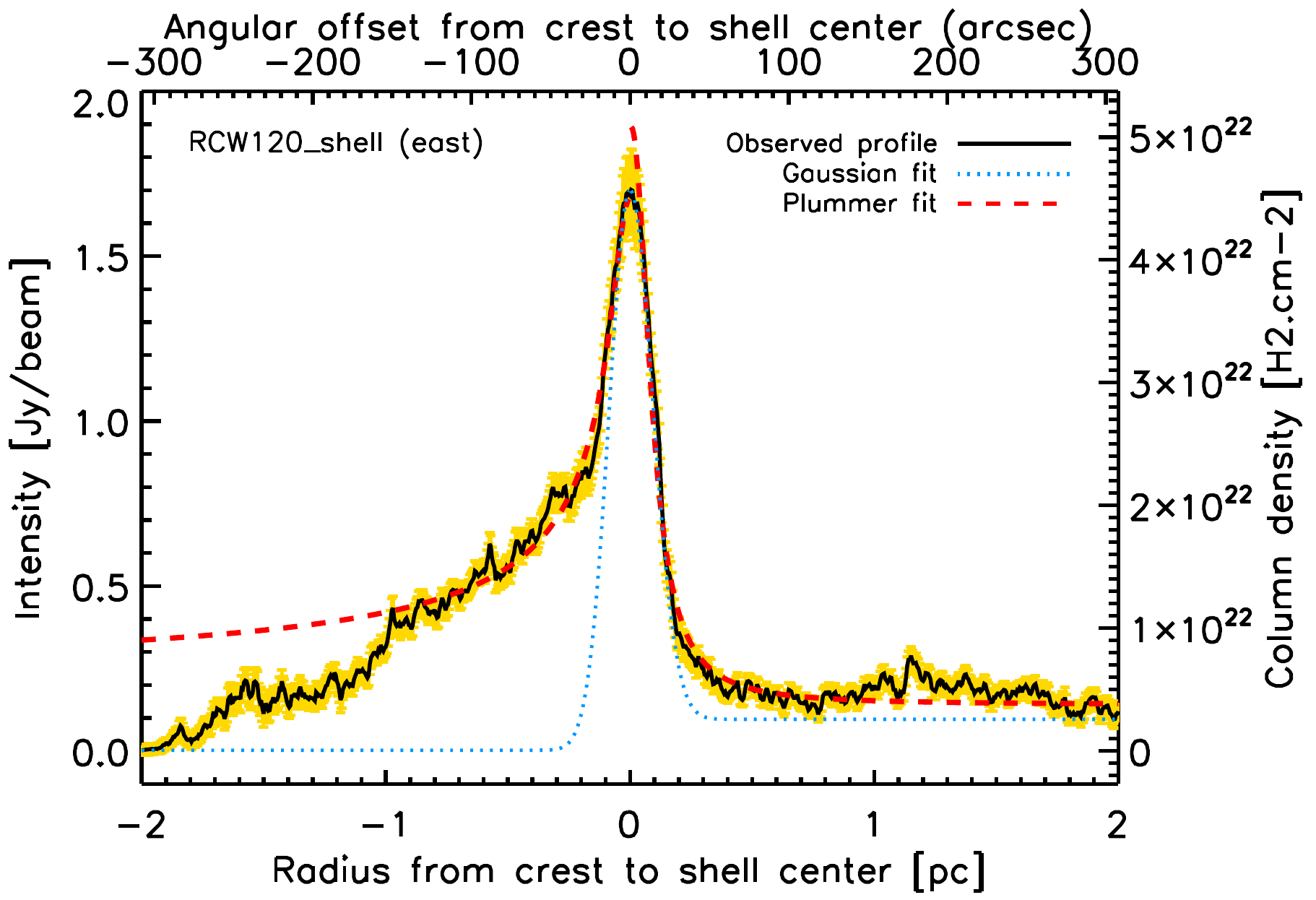}\hfill
\includegraphics[width=9cm]{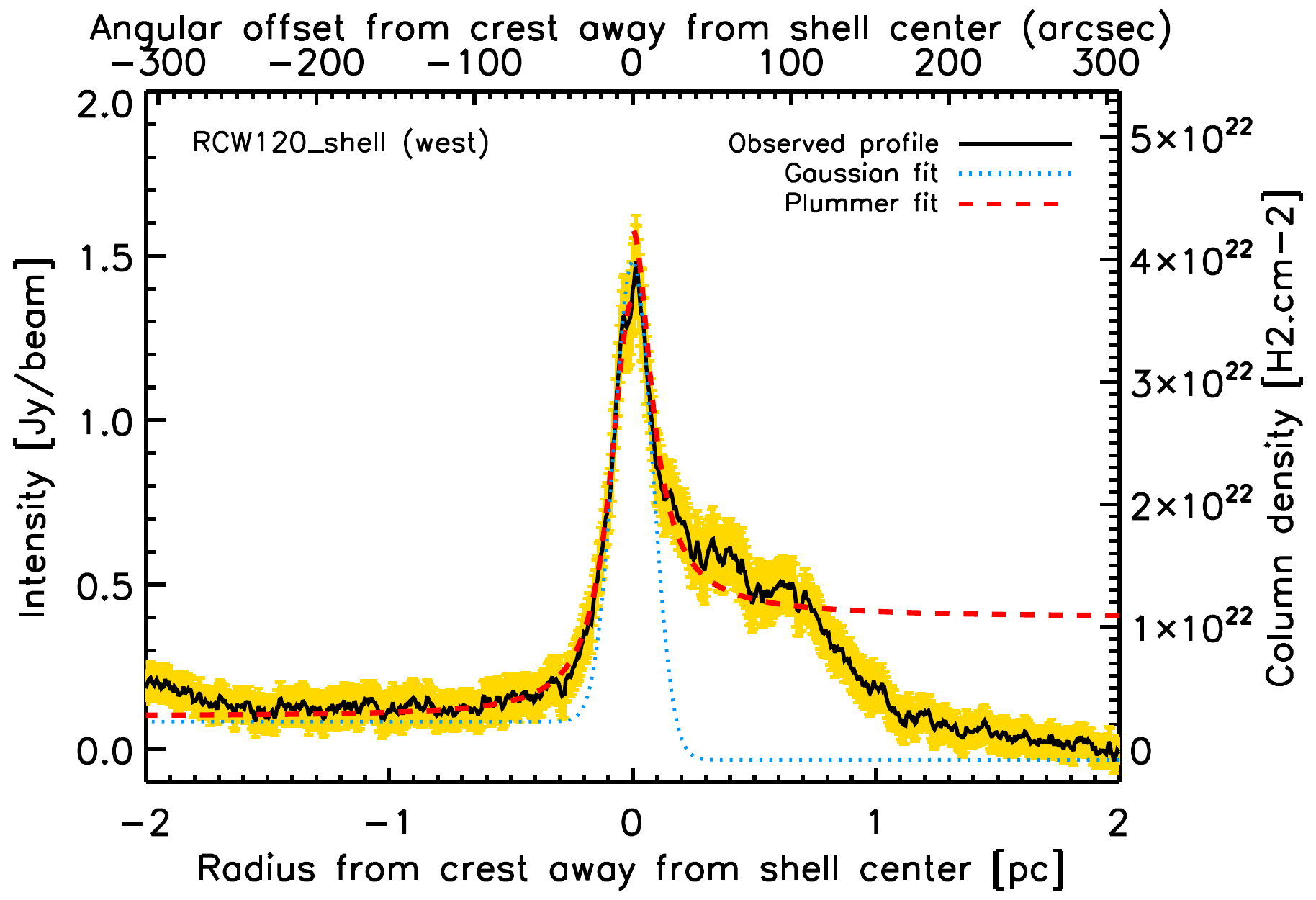}
   \caption{Mean radial intensity profile (black curve) of the eastern (left) and western (right) parts of the RCW~120 dense shell. The Gaussian fit (blue dotted curve) and the Plummer fit (red dashed curve) are shown. }
              \label{latradpro}%
    \end{figure*}

In order to study the variation compression over the dense shell, we studied the intensity profiles of the two lateral sides of the dense shell, shown as thick red lines in Fig.~\ref{fs}. These eastern and western parts of the resulting median radial intensity profile are displayed in linear format in Fig.~\ref{latradpro}. These profiles show that the two lateral sides of the shell are also asymmetric, as observed for the dense shell. 
This result is important because it indicates a homogeneous compression over the dense shell. 

\subsection{Profile of part of a radial filament} 
\begin{figure*}[h!]
\includegraphics[width=9cm]{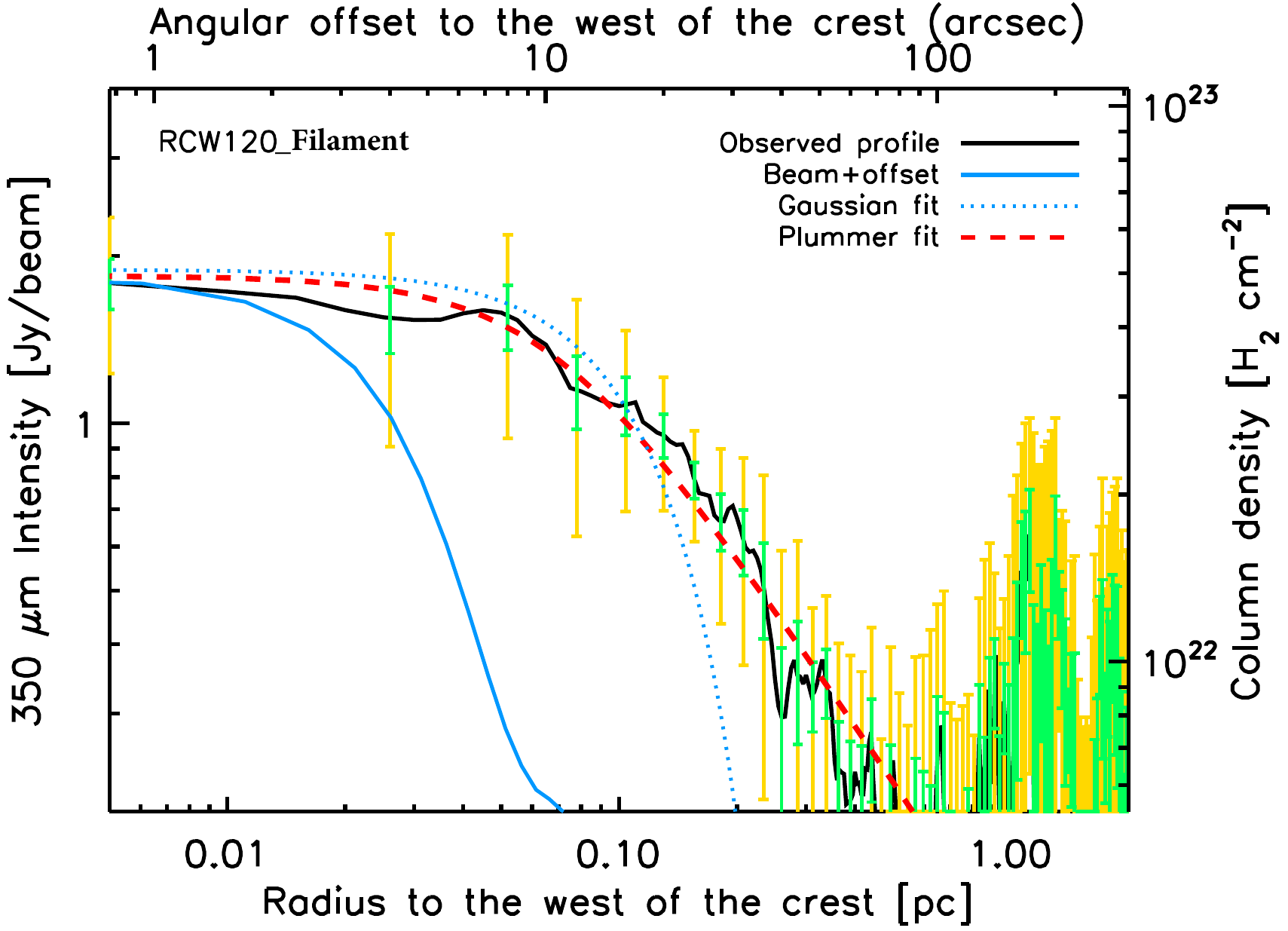}\hfill
\includegraphics[width=9cm]{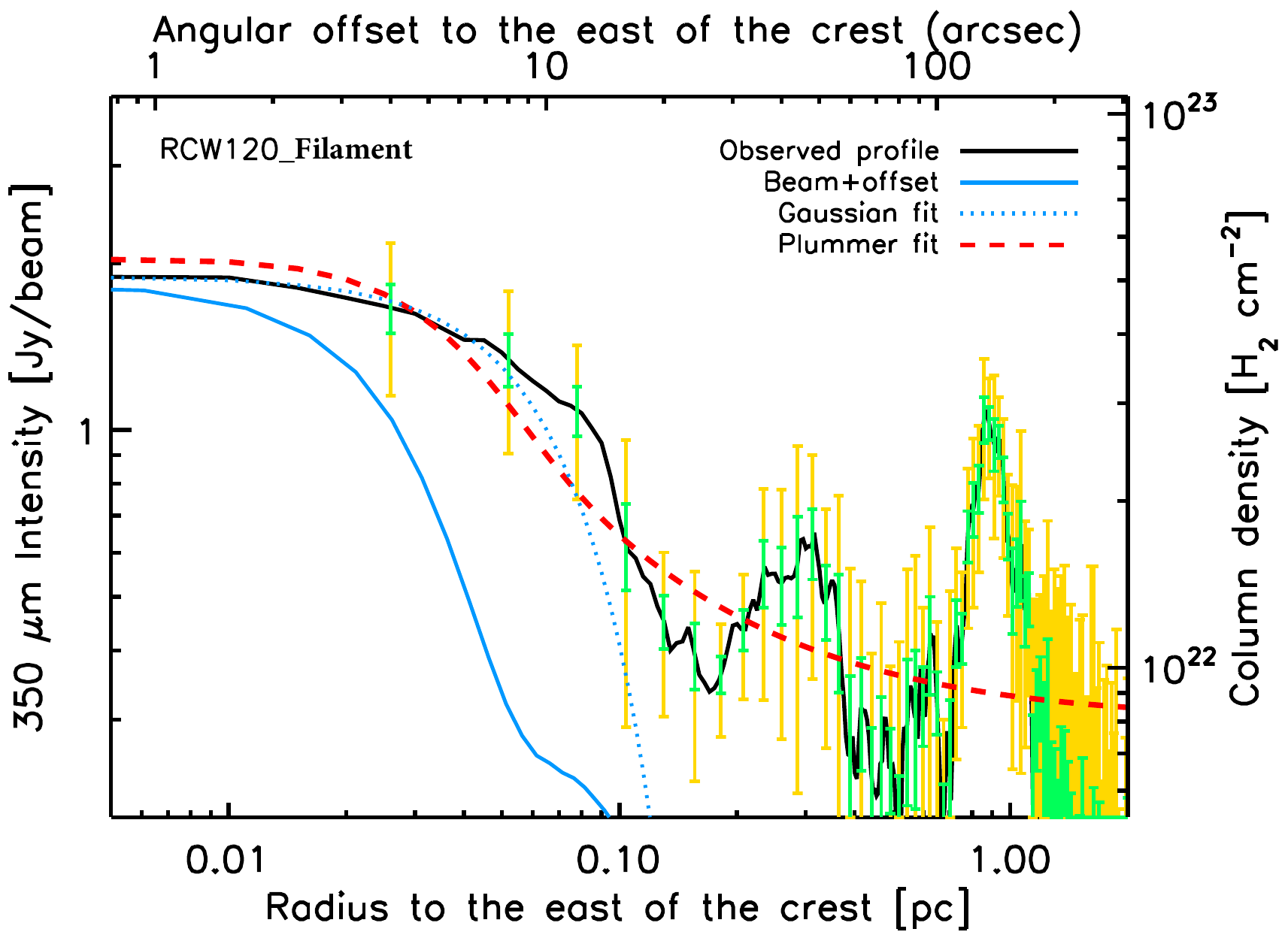}
   \caption{RCW~120: Mean radial intensity profile (black curve) of part of the radial filament, which is outlined in purple in Fig.~\ref{fs} and was measured perpendicular to the filament's crest in the western part (left) and in the eastern part (right). The green error bars show the standard deviation of the mean intensity profile. The blue solid curve shows the effective beam profile of the ArT\'eMiS 350 \mm data as measured on Mars. The blue dotted curve shows the best-fit Gaussian model to the observed profiles. The red dashed curve shows the best-fit Plummer model convolved with the beam. }
              \label{radprofil}%
    \end{figure*}

\begin{figure}[h!]
   \centering
  \includegraphics[scale=0.5]{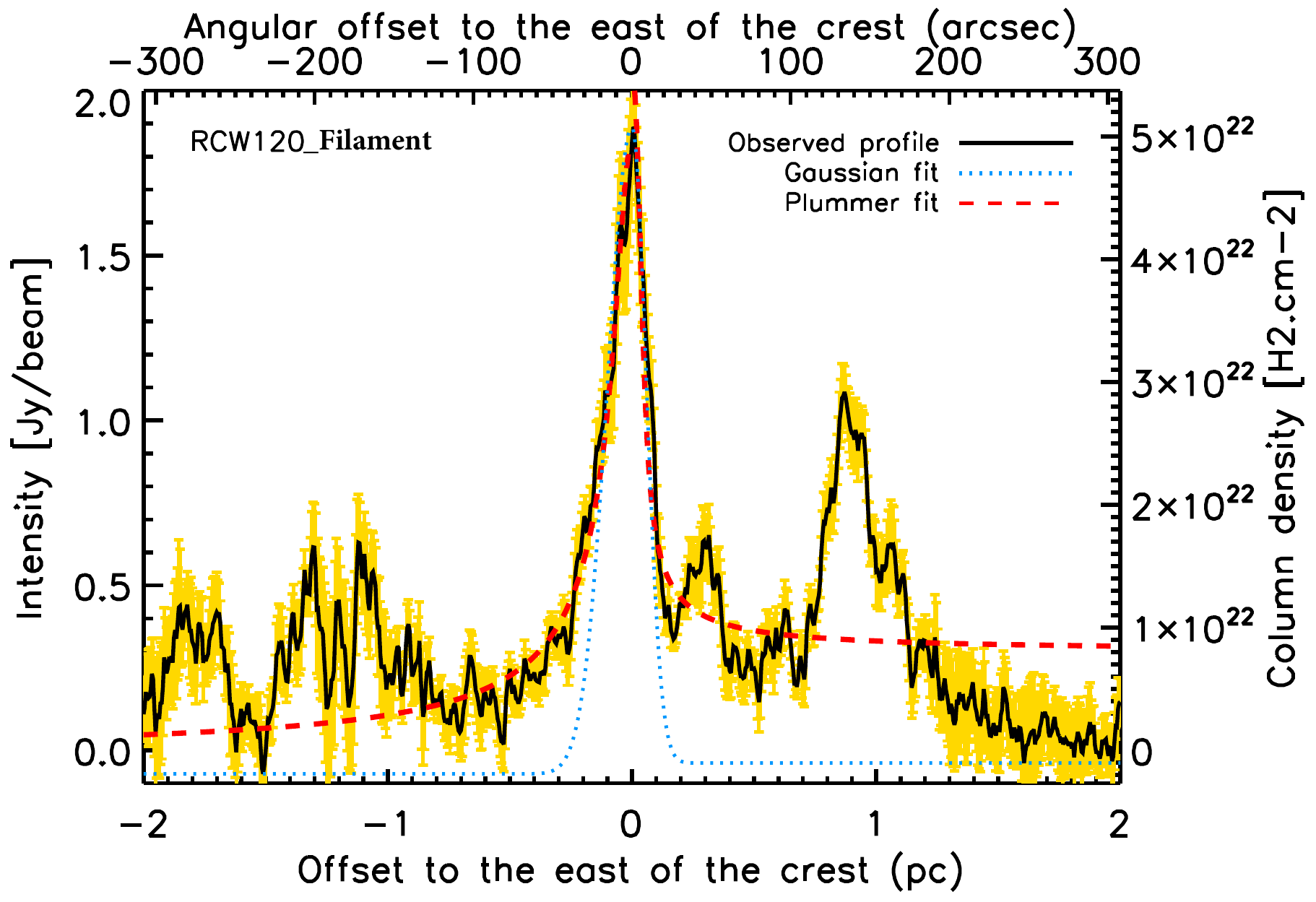}
   \caption{RCW~120: Transverse column density profile of the radial filament outlined in purple in Fig.~\ref{fs} (black curve). The Gaussian fit (blue dotted curve) and the Plummer fit (red dashed curve) are shown. Positive offsets are to the east. Emissions occuring out of the filament are also seen along the radial profile as secondary peaks. }
              \label{radprofillin}%
    \end{figure}

In order to study the properties of a radial filament that are connected to the shell but situated outside of it, we analyzed the profile of part of a radial filament of \sous. We selected this part of the filament, which is located in the southwest and shown in purple in Fig.~\ref{fs}, because it is bright enough to allow for this analysis. By taking perpendicular cuts at each pixel, we constructed two median radial intensity profiles. Fig.~\ref{radprofil} shows the shape of the radial intensity profile toward the western and eastern parts of this filament.  The profiles are noisier than the ones derived for the dense shell. The secondary emission peaks observed along the radial profile correspond to emissions observed along the cut, which are out of the filament. Fig.~\ref{radprofillin} shows the linear profile of a mean cut through the radial filament; the positive offset is to the east. This profile clearly shows a symmetric shape,  which is suggestive of no compression from the surroundings, contrary to what is observed for the shell (see Fig.~\ref{fa3}). 

\section{Discussion}\label{dis}
We are interested in the role of ionized regions in the formation of filaments. If \HII regions impact the formation of filaments, they also probably modify the properties of the sources the filaments host, that is, the cores and the stars that form inside. Therefore, it is important to study the relation between \HII regions and the filaments they impact. 
\subsection{Compression by the \HII region}
Using the \ar data presented here combined with \emph{Herschel} data, we revealed the structure of the dense shell observed toward \sous. The filament tracing the dense shell shows a clear asymmetric profile, which is suggestive of a radiative compression from the internal part of the ionized region (see Fig.~\ref{fa3}). The compression appears to be homogeneous over the dense shell, as shown in Fig.~\ref{latradpro}, suggesting that the observed shape of the dense shell that surrounds \sou results from the expansion of the ionized region (see Sect.~\ref{cc}).  
The radiative impact of \HII regions on their surroundings has been studied observationally and theoretically. Using \emph{Herschel} data, \citet{tre14} studied the impact of the ionization compression on dense gas distribution and star formation in four Galactic ionized regions, including \sous. In fitting the probability distribution function (PDF) of the column density of cold dust located around the ionized gas, they show that the PDF of all clouds is fitted with two lognormal distributions and a power law tail for high column density parts. For \sous, they show that the condensations that host star formation and are located at the edge of the ionized gas have a steep compressed radial profile, which is sometimes recognizable in the flattening of the power-law tail \citep{sch15}. This feature has been suggested to be used as an unambiguous criterion to disentangle triggered from pre-existing star formation. Their analysis of \sou shows that the most massive condensation observed at the southwestern edge (condensation 1, see \citealt{zav10,fig17}) may indicate the role of ionization compression for its formation and its collapse. The \ar data presented here and the analysis of the dense shell's profile show that compression occurs at the edges of \sous. Asymmetry in the filaments' profile  in the Pipe Nebula has been reported by \citet{per12} who used \emph{Herschel} images to suggest the formation of filaments by large-scale compression, which likely originate from the winds of the Sco OB2 association. They revealed a bow-like edge for the filamentary structures, suggesting the influence of the compression flow. In \sou, we also note the convexity of filaments, with the curvature oriented toward the ionized radiation, which is clearly seen in Fig.~\ref{fs} and particularly marked in the southwest where high-mass star formation occurs \citep{fig18}. An asymmetry in the filaments’ profile has also been reported in Orion B \citep{sch13}. They show that external compression broadens the column density probability distribution functions, as observed by \citet{tre14} for \sous.  On the modeling side, \citet{bro19} investigated whether photoevaporation of the illuminated edge of a molecular cloud could explain the high pressures and pressure versus UV field correlation observed at the PDR surface where warm molecular tracers are observed. Using the 1D Hydra PDR code, they show that photoevaporation can produce high thermal pressures ($P/k_{\rm{B}} \sim 10^7 - 10^8$ K cm$^{-3}$) and the observed $P - G_0$ correlation, almost independently from the initial gas density. They also show that the photoevaporating PDR is preceded by a low velocity shock (a few km/s) propagating into the molecular cloud. The high pressure resulting from photoevaporation could play a key role in the compression of the molecular material accompagnied by an increase in the local density. The compression is clearly seen in \sou and is also accompagnied by an increase in density, as was also observed by \citet{mar19}. The differential column density radial distribution that they present also shows a sharp profile on the interior side of the bubble (see their Figure 7), which is suggestive of an internal compression responsible for the sharp density enhancement. \citet{deh09} presented an unsharp mask image, free of extended emission, which was obtained from the Spitzer-MIPSGAL frame at 24 \mm near condensation~1 (see their Figure 10). This image shows distortions of the ionized-dense PDR interface under the influence of the ionizing stars' radiation observed perpendicular to it and with the same convexity reported here for the filaments.  

In \sou, the mean intensity profiles of the shell, which were measured toward the shell's center (internal part) and external part, are presented in Fig.~\ref{radpro} and are both well fitted by a classical Plummer profile. The FWHM of the internal part is 0.09~pc, which is similar to the width found for filaments in low-mass star-forming regions \citep{arz11,arz19}. This similar width observed in different environments suggests a similar origin for the filaments' formation. \citet{fed16} studied the universality of filaments using magnetohydrodynamical numerical simulations including gravity, turbulence, magnetic fields, as well as jet and outflow feedback. They found that these simulations reproduce realistic filament properties and they also found a universal filament width of $0.1 \pm 0.02$ pc. They propose that this characteristic filament width is based on the sonic scale ($\lambda_{\rm{sonic}}$) of the molecular cloud turbulence. This supports the idea that the filament width is determined by the sonic scale and gives a possible explanation for its universality. Using ALMA data, \citet{fig18} have recently shown that fragmentation is limited toward the most massive condensation (condensation 1) in \sou and they suggest that this can be due to the presence of a higher level of turbulence in this region. If so, the compression and turbulence may also have played a key role in the formation of filaments that host the high-mass star formation observed there. Evidence for compression revealed by the asymmetric profile of \sous's shell allows one to possibly link, for the first time, filaments' properties and star formation at the edges of an \HII region. 

\subsection{Symmetric profile of a radial filament}
As shown in Fig.~\ref{radprofillin} and contrary to what is observed for the dense PDR shell, the analysis of part of the southwest radial filament's profile (see Fig.~\ref{fs}) shows a clear symmetry, which is indicative of no compression for this filament. This suggests that the radiative compression from the \HII region is limited to the dense PDR shell and does not extend in the surrounding medium. However, the radial filaments observed around \sou may result from the leaking of radiation coming from the ionized region, as suggested by \citet{deh09} (see their Figure 16). The leaking radiation may shape the surrounding material in this case but, because the filament is parallel to the incoming radiation, no compression occurs, contrary to what happens in the dense shell where the material is perpendicular to the incoming radiation. 

\subsection{Cloud-cloud collision}  \label{cc}
\citet{tor15} present $^{12}$CO, $^{13}$CO, and C$^{18}$O data of \sou and suggest that the formation of the \sous's ionizing star could result from a cloud-cloud collision. Following this hypothesis, the authors suggest that the ionized region does not result from an expanding shell. According to the proposed scenario (see their Figure~12), the observed PDR layer located inside the cavity at the interface between the \HII region and the molecular cloud can be interpreted as the cavity created in the larger cloud by the collision of a smaller cloud. The inner surface of the interaction zone is illuminated by the strong ultraviolet radiation after the birth of the ionizing star. In this scenario, the compression over the shell is expected to be higher in the interaction zone between the small and large cloud, where the ionizing star has formed. However, we observe a similar asymmetric profile over the dense shell and on its lateral eastern and western parts (thick red lines in Fig.~\ref{fs}), suggesting an isotropic compression over the shell due to the expansion of the ionized region. \citet{and15} used MOPRA CO data to study the morphology of the \sou ionized region and found no evidence for expansion of the molecular material associated with \sous. However, they point out that the lack of detected expansion is roughly in agreement with models for the time-evolution of an \HII region like \sous, and this is consistent with an expansion speed of $\lesssim$~1.5 km s$^{-1}$. Forbidden lines, such as [O{\sc{i}}] and  [C{\sc{ii}}], are well suited to reveal the complex dynamics of \HII regions \citep{sch18} and will help to shed light on the internal dynamics of \sou (Luisi et al., in prep.).

\subsection{3D morphology of RCW~120}
The 3D morphology of RCW~120 is still debated. \citet{bea10} present CO (3$-$2) maps of 43 Galactic \HII regions and conclude that the shock fronts driven by massive stars tend to create rings instead of spherical shells. They propose that this morphology arises because the host molecular clouds are oblate. They also suggest that such a morphology implies that expanding shock fronts are poorly bound by molecular gas and that cloud compression by these shocks may be limited. However, the \ar data of \sou clearly reveal an asymmetric profile of the dense shell. 

\begin{figure}[h!]
   \centering
  \includegraphics[scale=0.5]{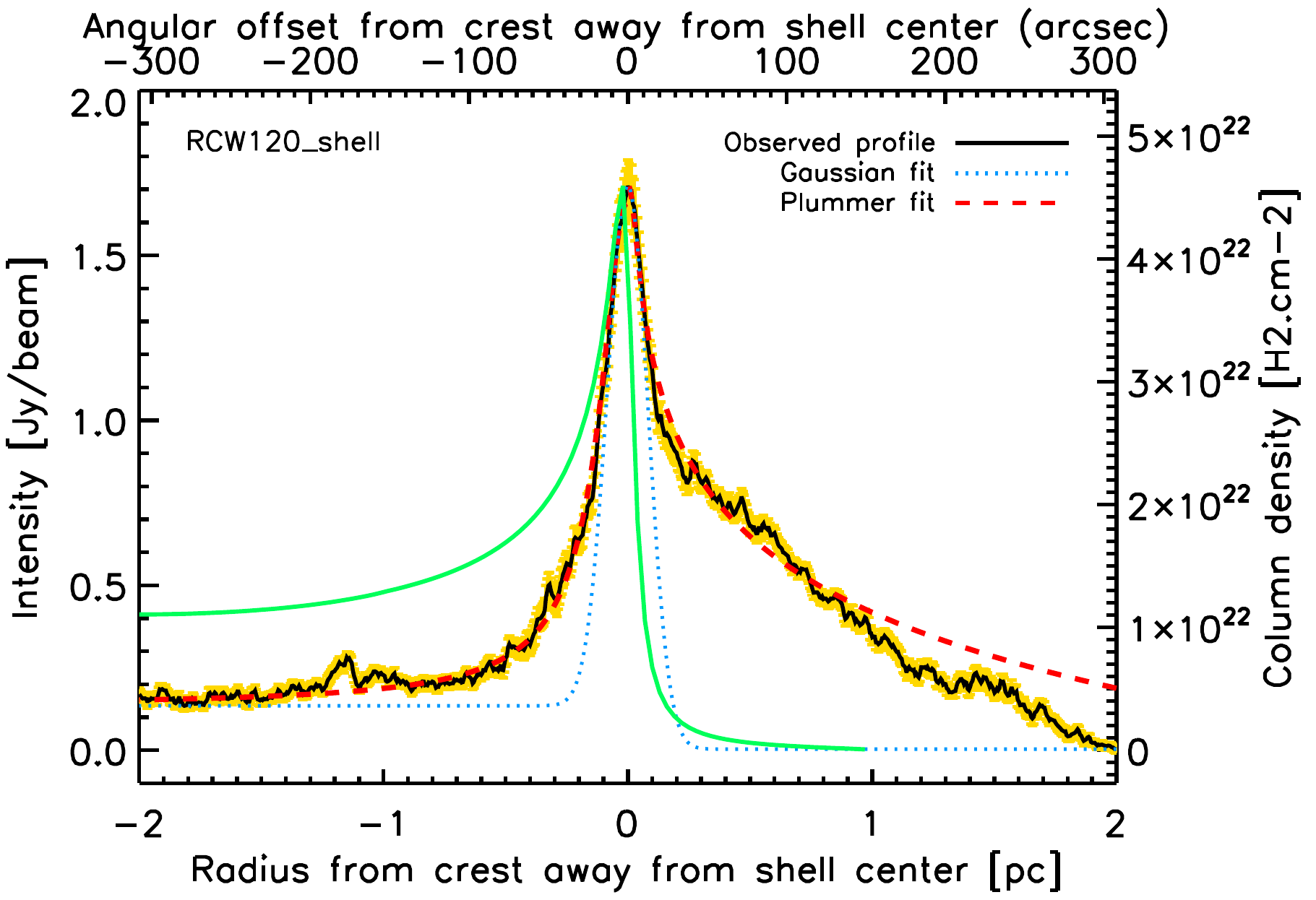}
   \caption{RCW~120: Observed mean radial column density profile (black curve) of RCW~120 dense shell, with positive offsets going away from the shell center. The Gaussian fit (blue dotted curve) and the Plummer fit (red dashed curve) are shown. The green full line shows the profile expected for a spherical shell (see text).}
              \label{morpho}%
    \end{figure}

Fig.~\ref{morpho} presents the observed median radial column density profile of the dense shell. The Gaussian fit (blue dotted curve) and the Plummer fit (red dashed curve) are shown, together with the profile expected for a spherical shell, assuming a classic Plummer profile fitted with R$_{Flat} \simeq  0.04$ pc and $p = 2$ (see Sect.~\ref{rps}). 
As observed in Fig.~\ref{morpho}, the model of a spherical shell produces too much emission toward the center of the region, compared with the observed profile. The model produces emission that is too low  compared to the observed emission of the shell's external part, but this could be due to extra emission from radial filaments observed in the emission map but which are not included in the model. To reproduce the observed emission, the model should have a very low R$_{Flat}$ and/or much higher $p$ values, which is in disagreement with the observations and results presented in Sect.~\ref{rps}. It is also important to note that the intensity level toward the shell's interior is slightly higher than that observed toward the exterior, indicating that material is present toward the center of the shell. This was also pointed by \citet{deh10} (see their Figure~12 and Sect.~6.1). Moreover, \citet{iwa11} present similar density profiles as the one observed here in their 3D simulations of gavitational fragmentation of expanding shells, pointing out the asymmetry of the density profile. The above arguments tend to favor an annular 3D morphology for \sous. This morphology is also favored by \citet{kir19} who modeled \sou with a ring-like face-on structure to reproduce the observed $^{13}$CO (2$-$1) and C$^{18}$O (2$-$1) lines (see also \citealt{pav13}). 

\section{Conclusions}\label{conc}
We have presented \ar 350 and 450 \mm emission maps toward the Galactic \HII\ region \sous, combined with $\emph{Herschel}$ data. We reveal the structure of the dense PDR associated with \sou and evidence for the compression of the internal \HII region from the asymmetric profile of the mean radial intensity profile of the dense shell. This compression is accompanied by a sharp increase in the column density and ongoing star formation. The width of the filament (0.09~pc) is similar to the one observed in low mass stars forming regions, pointing toward a universal formation mechanism. The analysis of the intensity profiles over the dense shell suggests that \sou could have an annular 3D morphology and that the homogeneous compression of the \HII region over the shell does not favor the cloud-cloud collision mechanism proposed for the formation of \sous's ionizing star.       

\begin{acknowledgements}
      AZ thanks the support of the Institut Universitaire de France. DA and PP acknowledge support from FCT through the research grants UIDB/04434/2020 and UIDP/04434/2020. PP receives support from fellowship SFRH/BPD/110176/2015 funded by FCT (Portugal) and POPH/FSE (EC).N.S. acknowledge support by the ANR/DFG grant GENESIS (ANR-16-CE92-0035-01/DFG1591/2-1). We thanks ESO, Onsala and the APEX staff in Chile for their support of the \ar project. We acknowledge the financial support from the French National Research Agency (Grants ANR–05–BLAN-0215 \& ANR–11– BS56–0010, and LabEx FOCUS ANR–11–LABX-0013), as well as supports from the French national programs on stellar and ISM physics (PNPS and PCMI). The research leading to these results has received funding from the European Union's Horizon 2020 research and innovation programme under grant agreement No 730562 [RadioNet]. Part of this work was also supported by the European Research Council under the European Union’s Seventh Framework Programme (ERC Advanced Grant Agreement No. 291294 – “ORISTARS”). This research has made use of data from the \emph{Herschel} HOBYS project (http://hobys-herschel.cea.fr). HOBYS is a  \emph{Herschel} Key Project jointly carried out by SPIRE Specialist Astronomy Group 3 (SAG3), scientists of the LAM laboratory in Marseille, and scientists of the  \emph{Herschel} Science Center (HSC).
\end{acknowledgements}

\bibliographystyle{aa} % style aa.bst
\bibliography{37815corr} 

\end{document}